\documentclass[
 aps,
 prb,
 amsmath,amssymb,
 reprint,
 floatfix,
 showpacs,
]{revtex4-1}

\usepackage{graphicx}
\usepackage{dcolumn}
\usepackage{bm}
\usepackage{textcomp}
\usepackage{hyperref}
\usepackage{verbatim}
\usepackage{enumitem}
\usepackage[usenames, dvipsnames]{color}
\usepackage{xfrac}
\usepackage[textsize=tiny,colorinlistoftodos]{todonotes}

\begin{document}

\preprint{APS/123-QED}

\title{Homodyne-detected ferromagnetic resonance of in-plane magnetized nano-contacts: composite spin wave resonances and their excitation mechanism}

\author{Masoumeh~Fazlali}
\affiliation{Department of Physics, University of Gothenburg, 412 96, Gothenburg, Sweden}
 
\author{Mykola~Dvornik}
\affiliation{Department of Physics, University of Gothenburg, 412 96, Gothenburg, Sweden}
 
\author{Ezio~Iacocca}
\altaffiliation{Current address: Department of Applied Mathematics, University of Colorado, Boulder, Colorado 80309-0526, USA}
\altaffiliation{Department of Applied Physics, Division for condensed matter theory, Chalmers University of Technology, 412 96, Gothenburg, Sweden}
\affiliation{Department of Physics, University of Gothenburg, 412 96, Gothenburg, Sweden}

\author{Philipp~D\"{u}rrenfeld}
\affiliation{Department of Physics, University of Gothenburg, 412 96, Gothenburg, Sweden}

\author{Mohammad Haidar}
\affiliation{Department of Physics, University of Gothenburg, 412 96, Gothenburg, Sweden}

\author{Johan~\AA{}kerman}
\affiliation{Department of Physics, University of Gothenburg, 412 96, Gothenburg, Sweden}
\affiliation{Materials Physics, School of ICT, KTH Royal Institute of Technology, 164 40 Kista, Sweden}
\affiliation{NanOsc AB, Electrum 205, 164 40. Kista, Sweden}

\author{Randy~K.~Dumas}
\email{randydumas@gmail.com}
\affiliation{Department of Physics, University of Gothenburg, 412 96, Gothenburg, Sweden}
\affiliation{NanOsc AB, Electrum 205, 164 40. Kista, Sweden}

\begin{abstract}
	This work provides a detailed investigation of the measured in-plane field-swept homodyne-detected ferromagnetic resonance (FMR) spectra of an extended Co/Cu/NiFe pseudo spin valve stack using a nanocontact (NC) geometry.  The magnetodynamics are generated by a pulse-modulated microwave current and the resulting rectified \emph{dc} mixing voltage, which appears across the NC at resonance, is detected using a lock-in amplifier.  Most notably, we find that the measured spectra of the NiFe layer are composite in nature and highly asymmetric, consistent with the broadband excitation of multiple modes. Additionally, the data must be fit with two Lorentzian functions in order to extract a reasonable value for the Gilbert damping of the NiFe. Aided by micromagnetic simulations, we conclude that \emph{(i)} for in-plane fields the \emph{rf} Oersted field in the vicinity of the NC plays the dominant role in generating the observed spectra,   \emph{(ii)} in addition to the  FMR mode, exchange dominated spin waves are also generated, and \emph{(iii)} the NC diameter sets the mean wavevector of the exchange dominated spin wave, in good agreement with the dispersion relation. 
\end{abstract}


\maketitle

\section{Introduction}

Spin torque ferromagnetic resonance (ST-FMR)\cite{Gao2005,Tulapurkar2005,Sankey2006,Sankey2008,Kubota2008,Wang2009,Tsoi2013,Wang2011,Mellnik2014} is a powerful and versatile tool that enables the characterization of magnetodynamics on the nanoscale. Unlike more conventional FMR measurement techniques, where a resonant cavity or waveguide is used to generate \emph{rf} magnetic excitation fields, the resonant precession in an ST-FMR measurement is assumed to be primarily a result of the ST from a spin polarized \emph{} current. However, ST-FMR represents a specific type of a more general homodyne detection scheme where the excitation mechanism itself can originate from a variety of physical mechanisms apart from, or in combination with ST, including, \emph{e.g.}, \emph{rf} Oersted fields \cite{Liu2011} and electric fields \cite{Nozaki2012}.

While homodyne-detected FMR studies on magnetic tunnel junction (MTJ) and all-metallic spin valve nano-pillar devices have dominated the literature\cite{Tulapurkar2005,Sankey2006,Sankey2008,Kubota2008,Wang2009}, there have been an increasing number of studies utilizing point- and nanocontacts (NCs) on extended multilayer film stacks\cite{Tsoi2013,Tsoi2011,Tsoi2011(2),Tsoi2015,Tsoi2014}. The NC geometry is particularly promising for high-frequency spin torque oscillators (NC-STOs) \cite{Silva2008, Dumas2015,Bonetti2010,Demidov2010, Madami2011,Bonetti2012,Sani2013} and for the emerging field of ST-based magnonics, \cite{Neusser2009a, Kruglyak2010, Chumak2015, Dumas2014b, Bonetti2013a} where highly non-linear auto-oscillatory modes are utilized for operation.

In the NC geometry literature\cite{Tsoi2013,Tsoi2011,Tsoi2011(2),Tsoi2015,Tsoi2014,Korenivski2014}, the observed ST-FMR spectra of the NiFe-based free layers have been analyzed as a single resonance, despite a significant peak asymmetry hinting at additional contributions. The linewidth of this asymmetric peak has not been understood so far\cite{Tsoi2013}. The same studies also note that the typical field condition of an in-plane field aligning both magnetic layers in parallel should not result in any ST, calling into question the fundamental excitation mechanism of the observed spectra. This significant discrepancy has been tentatively explained as being caused by local misalignments due to sample imperfections. However, given how robust ST-FMR measurements are over sets of different devices, it is rather unsatisfactory to have to refer to unknown extrinsic factors for the ST-FMR technique to function. It appears that the \emph{rf} Oersted field generated by the injected microwave current into NC  could be at play\cite{Tsoi2014}. Therefore, both a better fundamental understanding of the linear spin wave (SW) modes in the NC geometry and of their excitation mechanism are therefore highly desirable.

In this work we show that the observed resonance spectrum in a NiFe NC-STO free layer is \emph{composite} in nature and can be described as a  sum of two distinct resonances with very different behavior and origin. Experimentally, the data must be fit with two Lorentzian functions in order to extract a reasonable value for the Gilbert damping of the FMR mode. From these fits, it is also clear that only the width of one of the resonances shows a damping-like linear dependence on the frequency, whereas the width of the other is mostly frequency independent. Aided by micromagnetic simulations, we conclude that \emph{(i)} the composite resonance is a sum of a  FMR mode and an exchange dominated spin wave mode, \emph{(ii)} the NC diameter sets the mean wavevector of the exchange dominated spin waves, in good agreement with the dispersion relation, and \emph{(iii)} for in-plane fields the \emph{rf} Oersted field, not ST, in the vicinity of the NC plays the dominant role in exciting the observed spectra. We argue that homodyne-detected FMR studies in the NC geometry must account for such additional excitations to accurately extract the fundamental magnetodynamical properties. 
\begin{figure}[b] 
	\centering \includegraphics[width=3.3in]{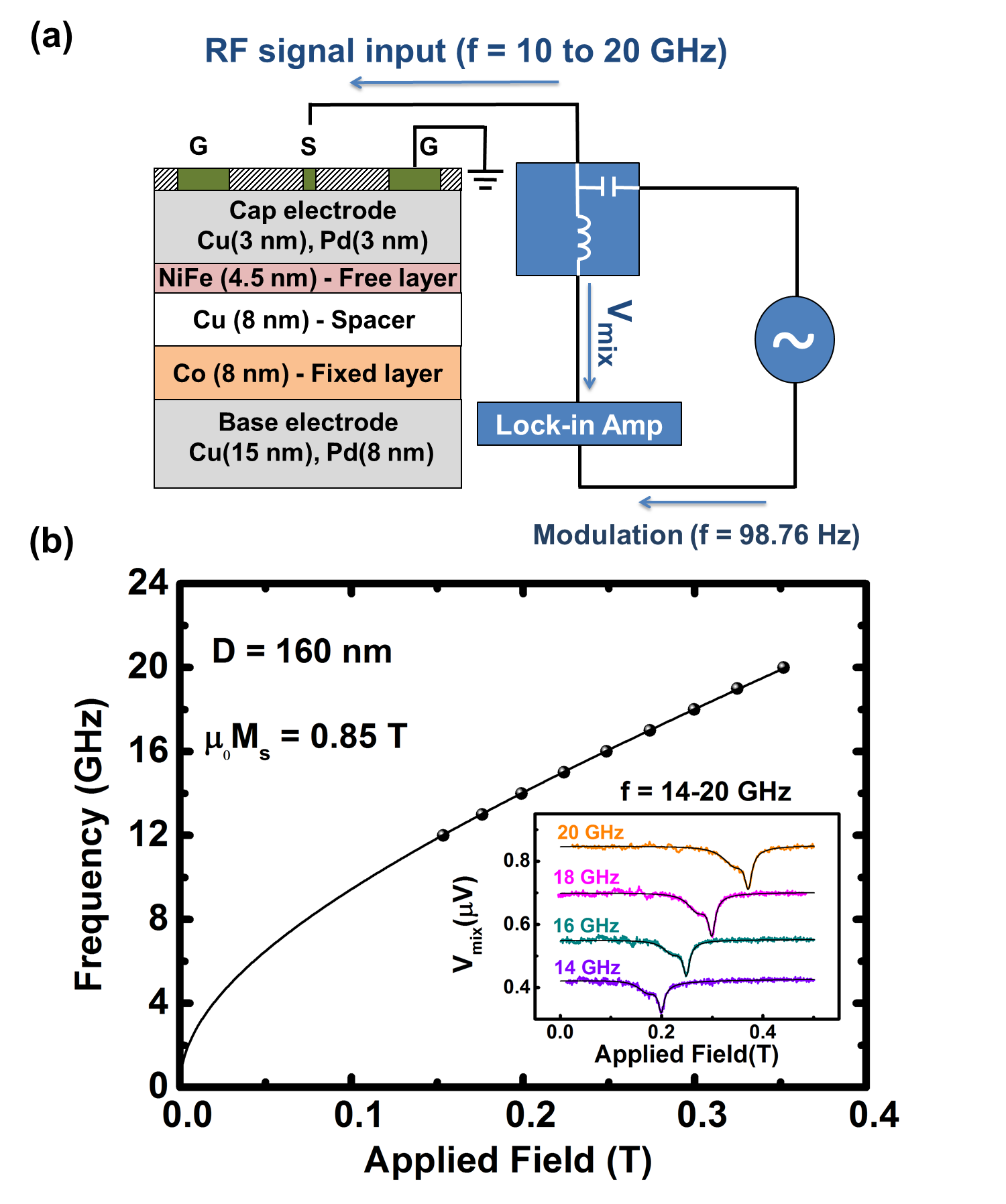} 
	\caption{
    (a) Schematic of the NC-STO and the measurement set-up. (b) Inset: ST-FMR spectra at four different frequencies for the $D$=160 nm sample. Main figure: Plot of the field position of the dominant resonance peak. The resonance fields can be well fit by the Kittel equation using $\mu_0M_s= $0.85 $\pm$ 0.02~T for the NiFe layer.}
	\label{fig:1}
\end{figure}

\section{Experiment}

NC-STO fabrication starts with a blanket Pd(8~nm) / Cu(15~nm) / Co(8~nm) / Cu(8~nm) / NiFe(4.5~nm) / Cu(3~nm) / Pd(3~nm) film stack deposited by magnetron sputtering on a thermally oxidized Si substrate, where the NiFe (Ni$_{80}$Fe$_{20}$) and Co play the role of the free and fixed layers, respectively, as shown in Fig.~\ref{fig:1}(a). The blanket film is then patterned into 16~$\mu$m x 8~$\mu$m spin valve mesas and a 30~nm SiO$_{2}$ layer is deposited by \textit{rf} magnetron sputtering. Circular NCs of nominal diameters, $D$, of 90~nm, 160~nm and 240~nm are defined through the SiO$_{2}$ insulating layer using e-beam lithography at the center of the mesa. A final photolithographic process then defines a coplanar waveguide for electrical connections and efficient microwave signal pick-up.  
\begin{figure}[b] 
	\centering \includegraphics[width=3.3in]{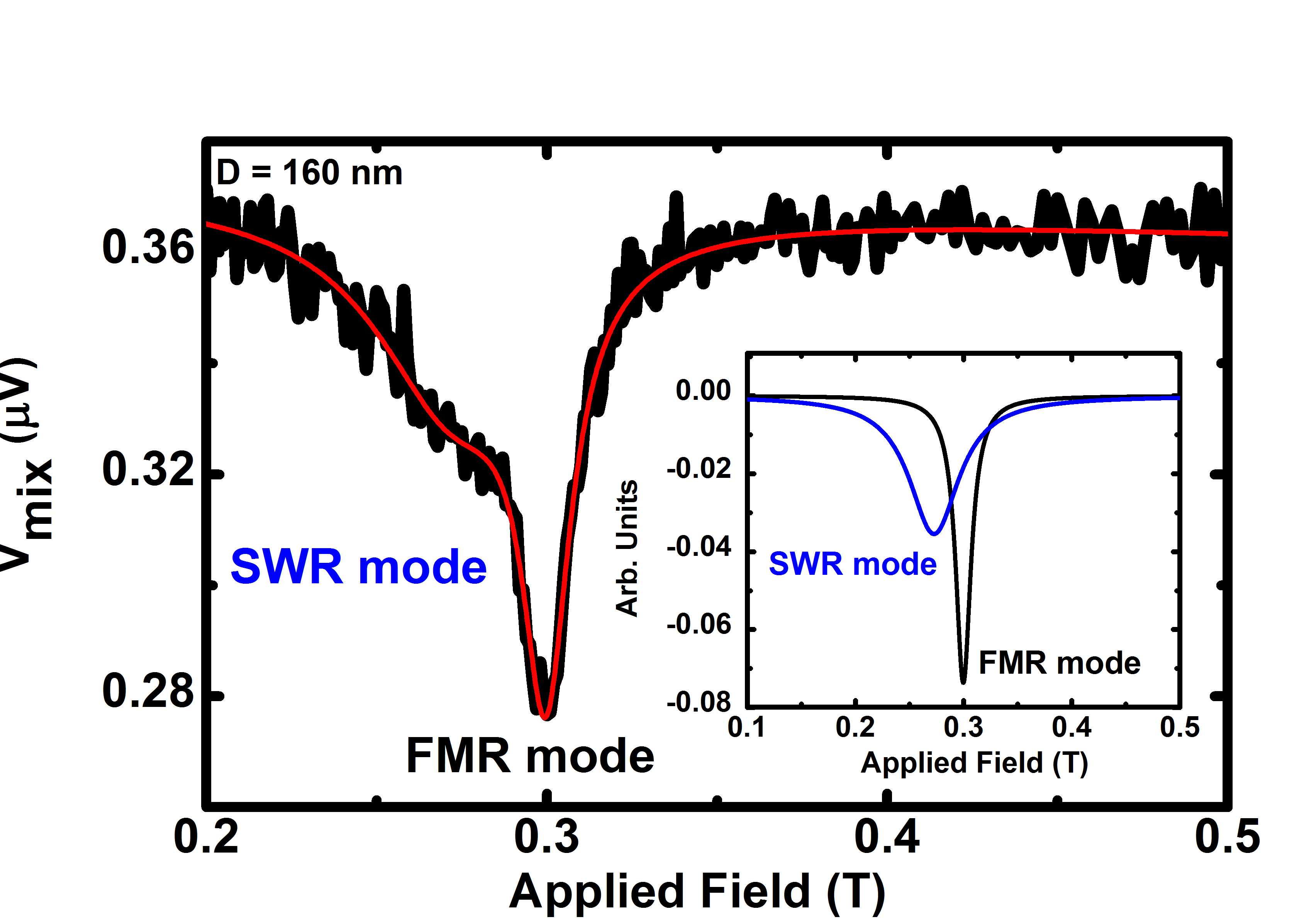} 
	\caption{Zoom-in of a representative ST-FMR spectrum of the $D$=160 nm sample taken at $f$ = 18~GHz and $I_{rf}$ = 1.3~mA, together with a fit (red line) based on two Lorentzians as described in the text. The inset shows the two individual contributions of the quasi-uniform FMR mode (black), and the spin wave resonance (blue).}
	\label{fig:2}
\end{figure}
All measurements were performed at room temperature in a custom built probe station utilizing a uniform \emph{in-plane} magnetic field. Our homodyne-detected FMR measurements utilized both a microwave generator and lock-in amplifier, which were connected to the device using a bias-tee, as schematically shown in Fig.~\ref{fig:1}. The \textit{rf} power injected into the NC is -14~dBm, which ensures that the excited magnetodynamics are in the linear regime. The resulting \textit{dc} mixing voltage\cite{Sankey2006}, $V_{mix}$, is measured as a function of the magnetic field and at a fixed excitation frequency. The microwave current was amplitude modulated at a low (98.76~Hz) modulation frequency for lock-in detection of $V_{mix}$.

\begin{figure}[b] 
	\centering \includegraphics[width=3.3in]{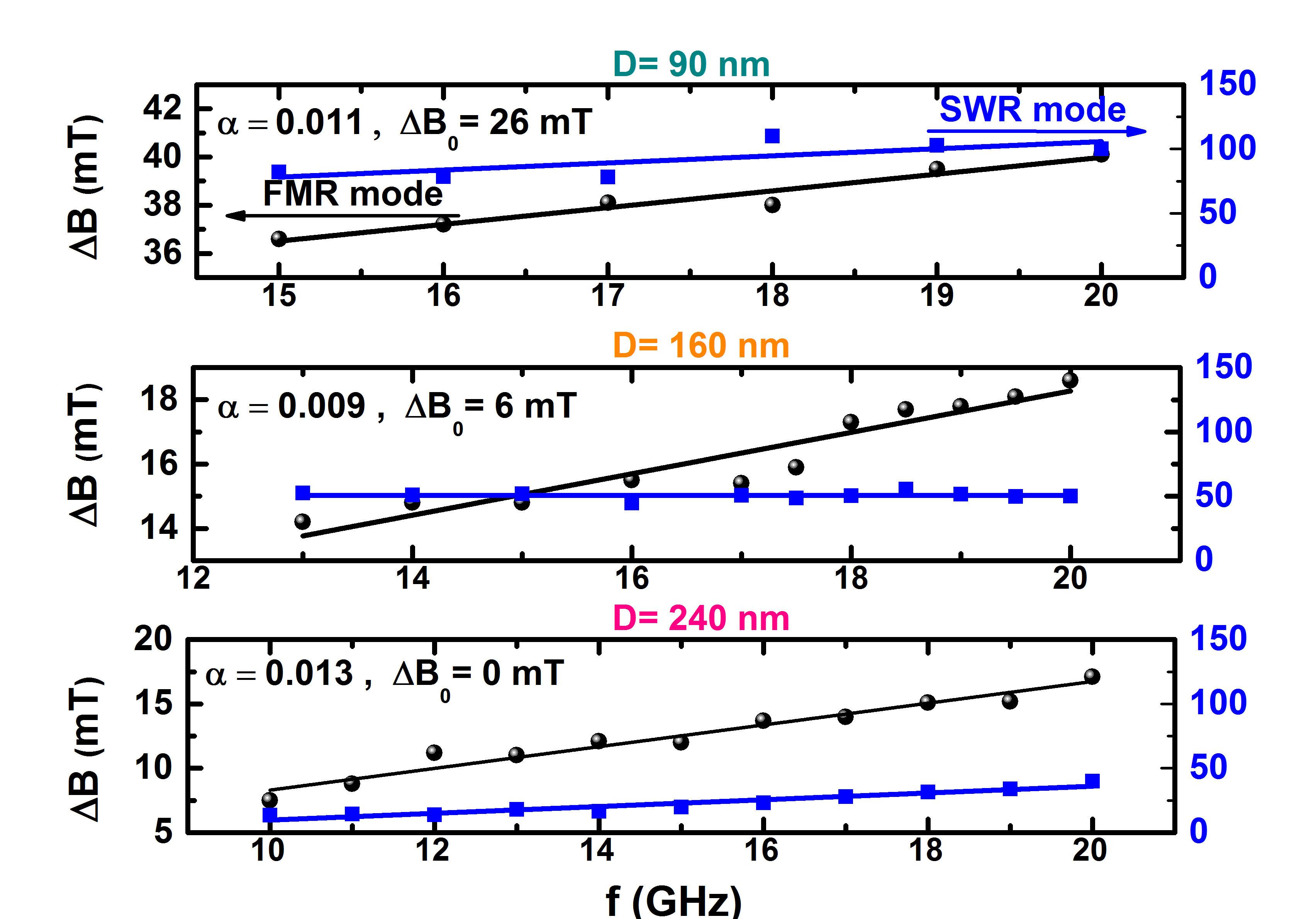} 
	\caption{The measured (dots) and fitted (solid lines) linewidths of the FMR and SWR modes are shown for the different NC diameters.}
	\label{fig:3}
\end{figure}

\section{Experimental Results}

The field-swept spectra measured for different frequencies, which are vertically offset for clarity, are shown in Fig.~\ref{fig:1}(b, inset) for the $D$=160 nm sample. As shown in the main panel of Fig.~\ref{fig:1}(b), the dominant resonance peak (data points) can be well fit (solid line) with the Kittel equation, which results in $\mu_0M_s=$ $0.85~\pm~0.02$~T and a negligible magnetocrystalline anisotropy.

Interestingly, upon closer inspection it becomes clear that the measured spectra are highly asymmetric, exhibiting a significant shoulder on the low-field side of the dominant resonance peak.  In Fig.~\ref{fig:2} we show a single representative resonance at $f$ = 18 GHz for the $D$=160 nm sample. While it is well known that the mixing voltage can be intrinsically asymmetric\cite{shafighadam2006,Tulapurkar2005,Kovalev2007}, it is important to point out that we cannot fit our data with a single resonance having both symmetric and antisymmetric contributions. Most importantly, the prior theoretical results are virtually independent of the NC diameter, in direct contrast to our experimental observations. In order to properly fit (red solid line) the entire spectrum we must instead use \emph{two} Lorentzian functions, each with its own resonance field and linewidth, as shown in Fig.~\ref{fig:2}(inset). The fit shows vanishing antisymmetric contribution to the lineshape for both of the resonances

\begin{equation}
\label{Equation:1}
\begin{split}
f&=\text{offset}+\sum_{i=FMR,SWR}\text{slope}\times B\\
&+\frac{1}{\Delta{B^{i}}}\bigg[S^{i}\frac{\Delta{B^{i}}^2}{4\left(B-B^{i}_{res}\right)^2+\Delta{B^{i}}^2}\\
&+A^{i}\frac{\Delta{B^{i}}(B-B^{i}_{res})}{4\left(B-B^{i}{res}\right)^2+\Delta{B^{i}}^2}\bigg]\text{,}
\end{split}
\end{equation}
where $B$ and $B^{i}_{res}$ are applied and resonance fields, respectively, and $\Delta B^{i}$ is the linewidth of the corresponding peak. $S^{i}$ and $A^{i}$ are amplitudes of its symmetric and anti-symmetric components, respectively. As the frequency vs.~field behavior of the main resonance mode can be well fit with the Kittel equation, Fig.~\ref{fig:1}(b), we ascribe this peak to the  FMR mode of the NiFe layer and the second low field mode with a higher order spin wave resonance (SWR), which will be discussed in detail later.  

The linewidth vs.~frequency of both the FMR and SWR modes are plotted in Fig.~\ref{fig:3} for three different NC diameters of 90, 160, and 240~nm. Three different significant observations can be made. First, the FMR mode shows a clear linear increase of linewidth with the frequency, from which the Gilbert damping, $\alpha$, can be extracted using the following relation:
\begin{equation}
	\label{Equation:2}
	\Delta B^{i} = \frac{4\pi\alpha}{\gamma}f +\Delta B^{i}_{0}
\end{equation}
where $\Delta B^{i}_0$ is an inhomogeneous broadening of the corresponding resonance. Our measured values of $\alpha$, which are all on the order of 0.01, are also consistent with those measured in Ref.~\onlinecite{Houshang2014}. This provides further evidence that the dominant resonance mode can indeed be correlated with the usual FMR mode of NiFe. Second, the linewidth of the SWR mode is mostly independent of frequency, indicating that the primary origin of the linewidth is not damping. Third, the inhomogeneous broadening is approximately \emph{inversely proportional} to the NC diameter, which at first seems counter intuitive as one would expect a larger NC to sample a larger sample volume and therefore include more inhomogeneities. The origin of this interesting effect will be explained later.

\begin{figure} 
	\centering \includegraphics[width=3.3in]{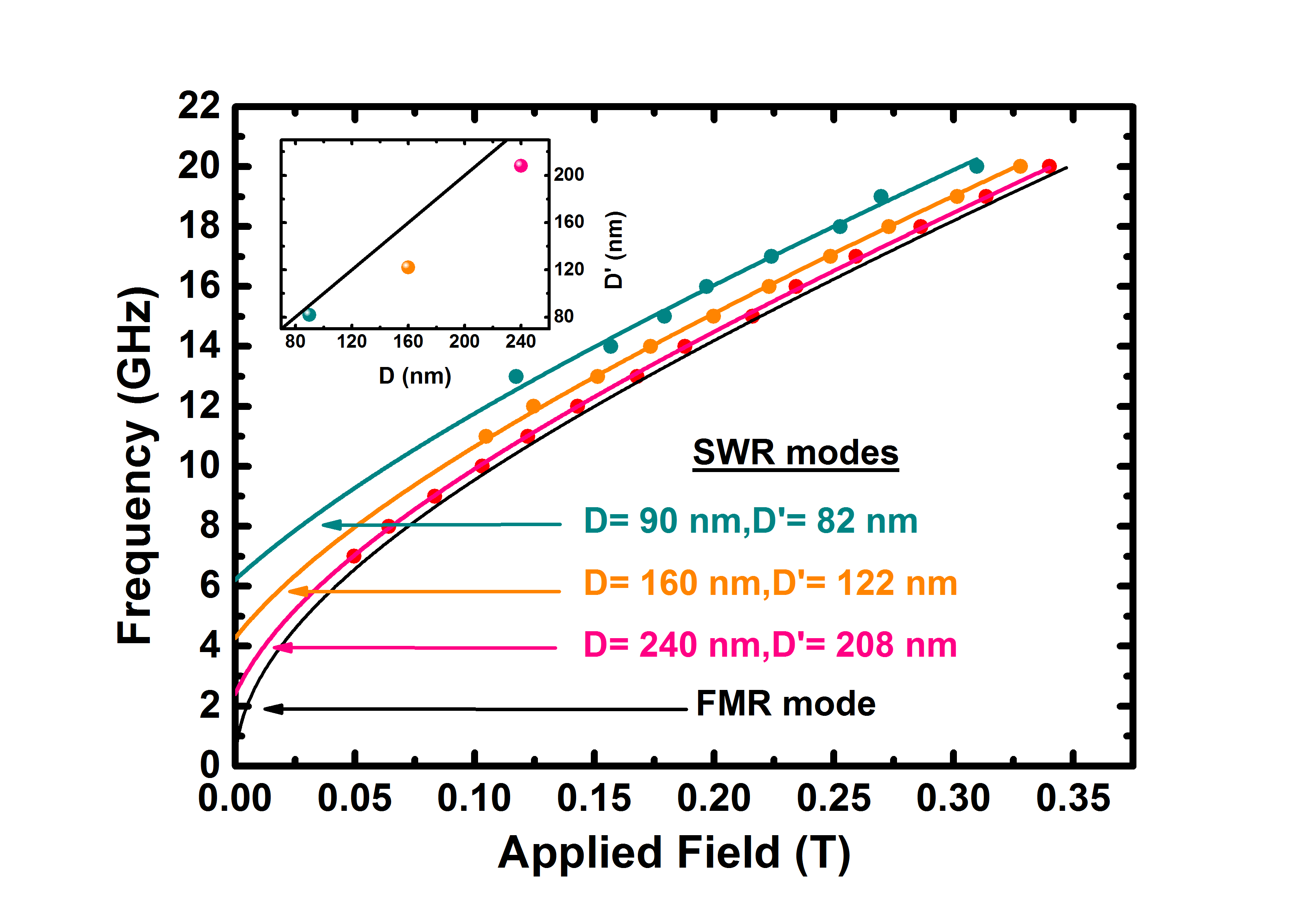} 
	\caption{Measured (dots) and calculated (solid lines) resonance fields of the FMR and SWR modes for the different NC diameters. The black solid line is a fit to an average of the the FMR mode for all three devices. Inset: A plot of the fitted NC diameter ($D^{\prime}$) vs.~the nominal diameter ($D$), together with a line indicating $D^{\prime}=D$. 
    \label{fig:4}}   
\end{figure}

The frequency versus field dependence of the measured FMR and SWR modes are summarized in Fig.~\ref{fig:4}.  The black solid line shows the average behavior of the FMR mode for NC diameters of 90, 160, and 240~nm and essentially reproduces what is shown in Fig.~\ref{fig:1}. For a fixed frequency, we find that the SWR mode shifts to lower fields as the NC diameter decreases. Assuming that the origin of the SWR mode is the exchange interaction, the diameter of the NC, $D^{\prime}$, can be estimated using the following dispersion relation:
\begin{eqnarray}
	\label{Equation:3}
	f &=& \frac{\gamma}{2\pi}\bigg[\left(B^{SWR}_{res}+\mu_0M_{S}\left(\lambda_{ex}k\right)^2\right)\nonumber\\
	&&\times\left(B^{SWR}_{res}+\mu_0M_{S}+\mu_0M_{S}\left(\lambda_{ex}k\right)^2\right)\bigg]^{1/2}{,}  
\end{eqnarray}
where $\lambda_{ex} = \sqrt{2A/\mu_0M_s^2}$  and $k=\pi/D^{\prime}$ are the exchange length and the SWR wave vector, respectively. The room temperature value of the exchange stiffness is set to $A=11$~pJ/m.\cite{Yin2015}  The estimated sizes of the NCs are in reasonable agreement with the corresponding nominal values as shown in the inset of Fig.~\ref{fig:4}.

\section{Micromagnetic Simulations}

The micromagnetic simulations were performed using the mumax3 solver~\cite{Vansteenkiste2014}. Since the actual spin valve mesa is too large to be simulated in its entirety in a reasonable time frame, we limited our calculations to a 5.120~$\mu$m $\times$ 2.560~$\mu$m $\times$ 4~nm volume with periodic boundary conditions tailored to mimic the lateral aspect ratio of the experimental spin valve mesa. To break the symmetry of the system, which might otherwise fully forbid any STT-related effects and non-conservative SW scattering, we assume that the applied field points $5$\textdegree out-of-plane, comparable to the possible error in the experimental field alignment. As a first step, the evolution of the ground state of the entire Co/Cu/NiFe stack is calculated, confirming that \emph{(i)} the Co and NiFe layers remain virtually collinear in the given range of the applied magnetic fields and \emph{(ii)} there are no mutual stray fields produced between the layers in the vicinity of the NC. Since there is a significant spin wave dispersion mismatch between Co and NiFe, we do not expect any resonant dynamic magnetic coupling between the layers. Under these three 
considerations we can confidently simulate the dynamics of the NiFe free layer alone. 

\begin{figure}
	\centering \includegraphics[width=3.3in]{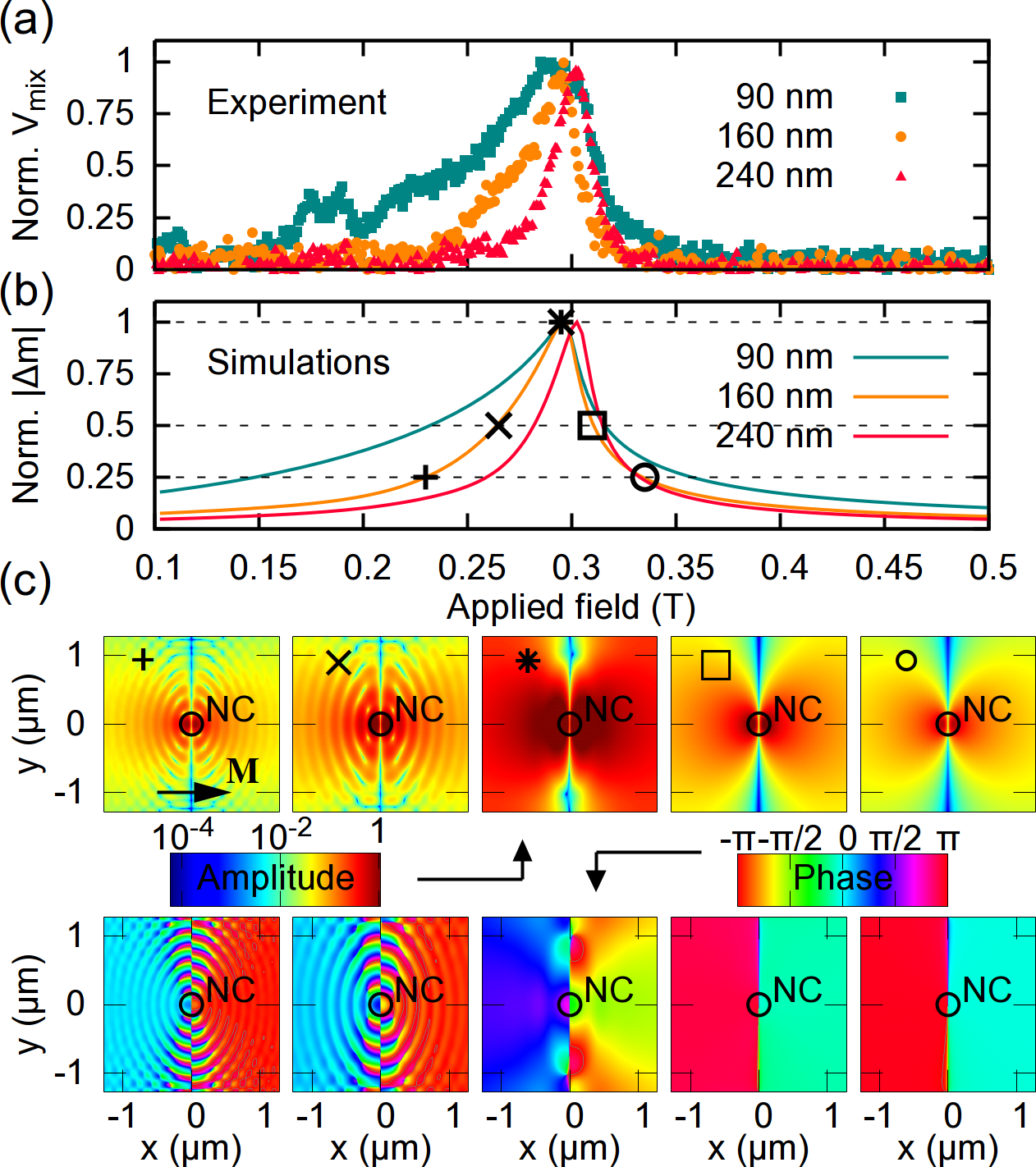} 
	\caption{(a) The normalized measured mixing voltage ($V_{mix}$) and (b) the normalized simulated magnetization precesssion amplitude for the three NC diameters as a function  of the applied in-plane magnetic field. (c) Spatial maps of the magnetization precession amplitude (top row) and phase (bottom row) simulated for a $D = 160$ nm NC diameter taken at the fields corresponding to the main peak and its \sfrac{1}{2} and \sfrac{1}{4} heights (as shown by the corresponding black symbols in (b)). Propagating spin waves are clearly seen for the two lowest fields.}
	\label{fig:5}
\end{figure}

In the simulations we replicate the experimental data acquisition routine by performing the field sweeps with a harmonic excitation of $f=18$ GHz. The infinite wire approximation is used to calculate the Oersted field produced by the NC.\cite{Petit-Watelot, Randy2013} For every value of the applied field we first let the system reach the steady state and then sample the spatial map of the magnetization for the following 5~ns at 3.5~ps time intervals with a subsequent point-wise FFT applied and the amplitude and phase of the magnetization precession extracted at the excitation frequency. Where applicable, the direction of the spin-current polarization is assumed to be collinear with the magnetization in the nominally fixed Co layer. The implemented saturation magnetization, gyromagnetic ratio, and damping constant are estimated by fitting a Kittel equation to the experimental data. The room temperature value of the exchange stiffness is set to $A=11$~pJ/m.\cite{Yin2015}

The simulated magnetic response shown in Fig.~\ref{fig:5}(b) agrees well with the experimentally measured data shown in Fig.~\ref{fig:5}(a). To identify the origin of the observed peak asymmetry we investigate the spatial profiles of the magnetization precession amplitude in the vicinity of the resonance, see Fig.~\ref{fig:5}(c). The snapshots clearly show propagating SWs on the low-field side of the main peak, while no SWs are resolved on the high-field side. Looking closer at the phase profiles of the corresponding modes, that essentially depict the wavelength of the excited magnons, the following conclusions can be made: (a) The propagation of SWs perpendicular to the saturation direction is suppressed and (b) the lowest excited mode is not uniform, but anti-symmetric with respect to the NC center.

\section{Discussion}

\begin{figure}[b] 
	\centering \includegraphics[width=3.3in]{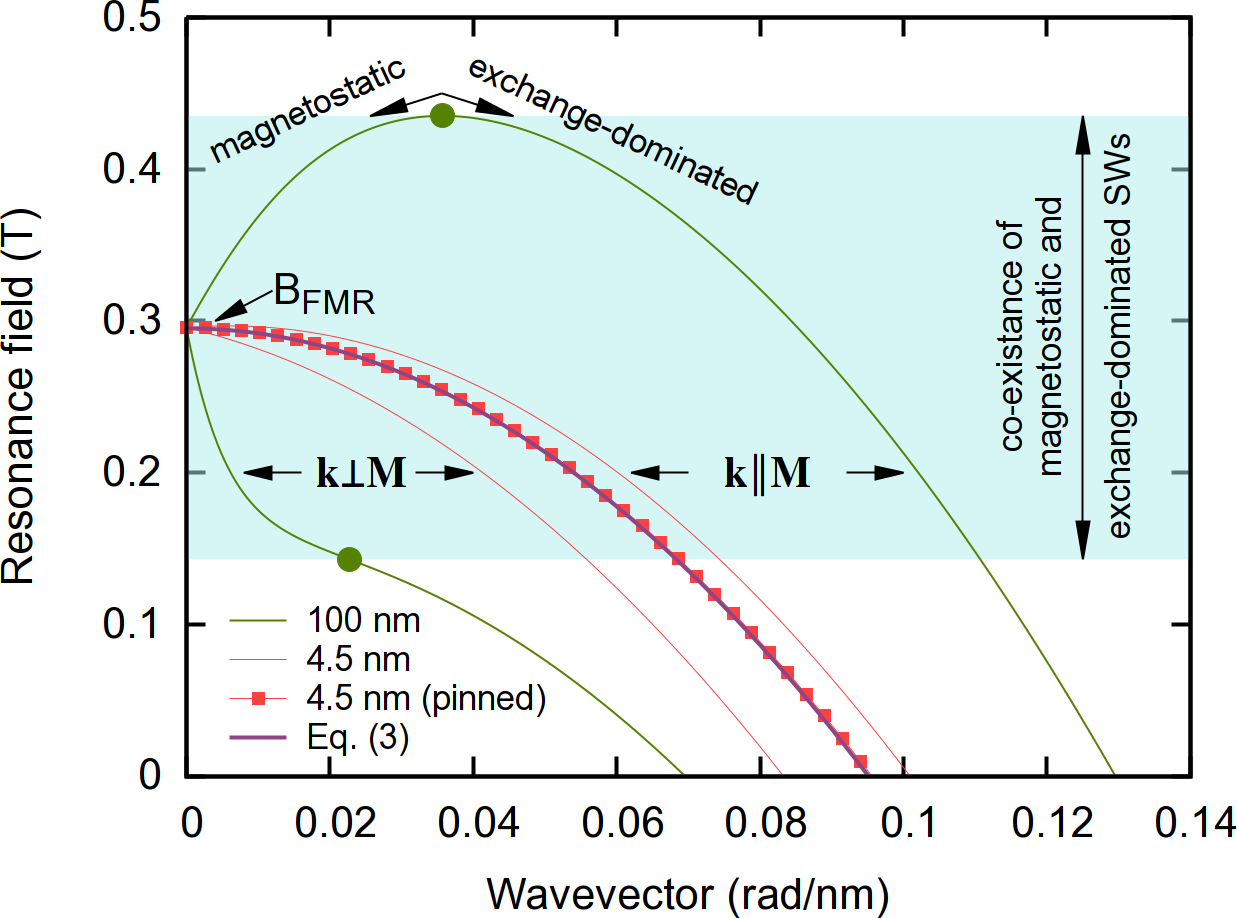} 
	\caption{The dispersion relations for the SWs propagating parallel and perpendicular to the saturation direction are shown for the different thicknesses of the NiFe layer. The points correspond to the minimum of the SW group velocity.
    \label{fig:6}}
\end{figure}

If the free layer is magnetized in-plane, then both backward volume magnetostatic SWs (BVMSSW) and surface magnetostatic-exchange SWs (SMSSW) can be excited: 
\begin{eqnarray}
	\label{Equation:BVMSSW}
	f_{BVMSSW} &=& \bigg[\left(f_B+f_M\left(\lambda_{ex}k\right)^2\right) \nonumber \\
	&&\times\left(f_B+f_M\left(\lambda_{ex}k\right)^2+f_M\left( \frac{1 - \operatorname{e}^{-kd}}{kd}\right)\right)\bigg]^{1/2}{,} \nonumber
\end{eqnarray}

\begin{eqnarray}
	\label{Equation:SMSSW}
	f_{SMSSW} &=& \bigg[\left(f_B+f_M\left(\lambda_{ex}k\right)^2\right) \nonumber \\
	&&\times\left(f_B+f_M\left(\lambda_{ex}k\right)^2+f_M\right) \nonumber \\
    &&+\frac{f_{M}^2}{4}\left( 1 - \operatorname{e}^{-2kd}\right)\bigg]^{1/2}{,} \nonumber
\end{eqnarray}

where $f_B = \frac{\gamma}{2\pi}B$, $f_M = \frac{\gamma}{2\pi}\mu_0M_{S}$

They are calculated using Eq.~(5.97b)  and Eq.~(5.111b) from Ref.~\onlinecite{Stancil2009} for propagation along and perpendicular to the saturation direction, respectively.  The exchange contribution is included by substituting $B\rightarrow B+\mu_0M_s\left(\lambda_{ex}k\right)^{2}$. 

The corresponding dispersion relations are shown in Fig.~\ref{fig:6} for Py thicknesses, $d$, of 100 nm (green lines) and 4.5 nm (red lines). There is always a region of resonance fields, where magnetostatic and exchange-dominated SWs co-exist, as highlighted by the shaded area in Fig.~\ref{fig:6} for the Py thickness of 100~nm.  Although the band is broad for relatively thick layers, it only amounts to 1.16~mT for the 4.5~nm Py, \emph{i.e.}, an order of magnitude smaller than the intrinsic broadening of the FMR peak. We therefore conclude that SWs contributing to the low-field tail of the FMR peak are exchange-dominated. Note that the calculated dispersion relations differ from what is found using Eq.~(\ref{Equation:3}) (thick solid line in Fig.~\ref{fig:6}).  This difference arises as the dispersion relations also strongly depend on the exact boundary conditions at the free layer surfaces. For instance if the Py film is pinned on both surfaces, \emph{e.g.}, if placed in between sufficiently thick metallic layers, the dispersion of the exchange-dominated backward volume SWs is given by the following equation (shown by the dotted line in Fig.~\ref{fig:6})~\cite{A.G.Gurevich1996}:
\begin{eqnarray}
	\label{Equation:pinned-bvmssw}
	f_{BVMSSW-pinned} &=& \bigg[\left(f_B+f_M\left(\lambda_{ex}k\right)^2\right) \nonumber \\
	&&\times\left(f_B+f_M\left(\lambda_{ex}k\right)^2+\frac{f_M}{1+(kd/\pi)^2}\right)\bigg]^{1/2}{,} \nonumber
\end{eqnarray}
Note that in this case the spectrum of the exchange-dominated surface SWs will be dispersionless and not accessible experimentally 

Since Eq.~(\ref{Equation:3}) fits the NC diameter reasonably well, we conclude that (a) the detected mixing voltage is generated by the exchange-dominated backward volume SWs and (b) there is undoubtedly some surface pinning of the Py layer. The exact origin of the pinning and its strength is beyond the scope of the present study.

\begin{figure}[b] 
	\centering \includegraphics[width=3.3in]{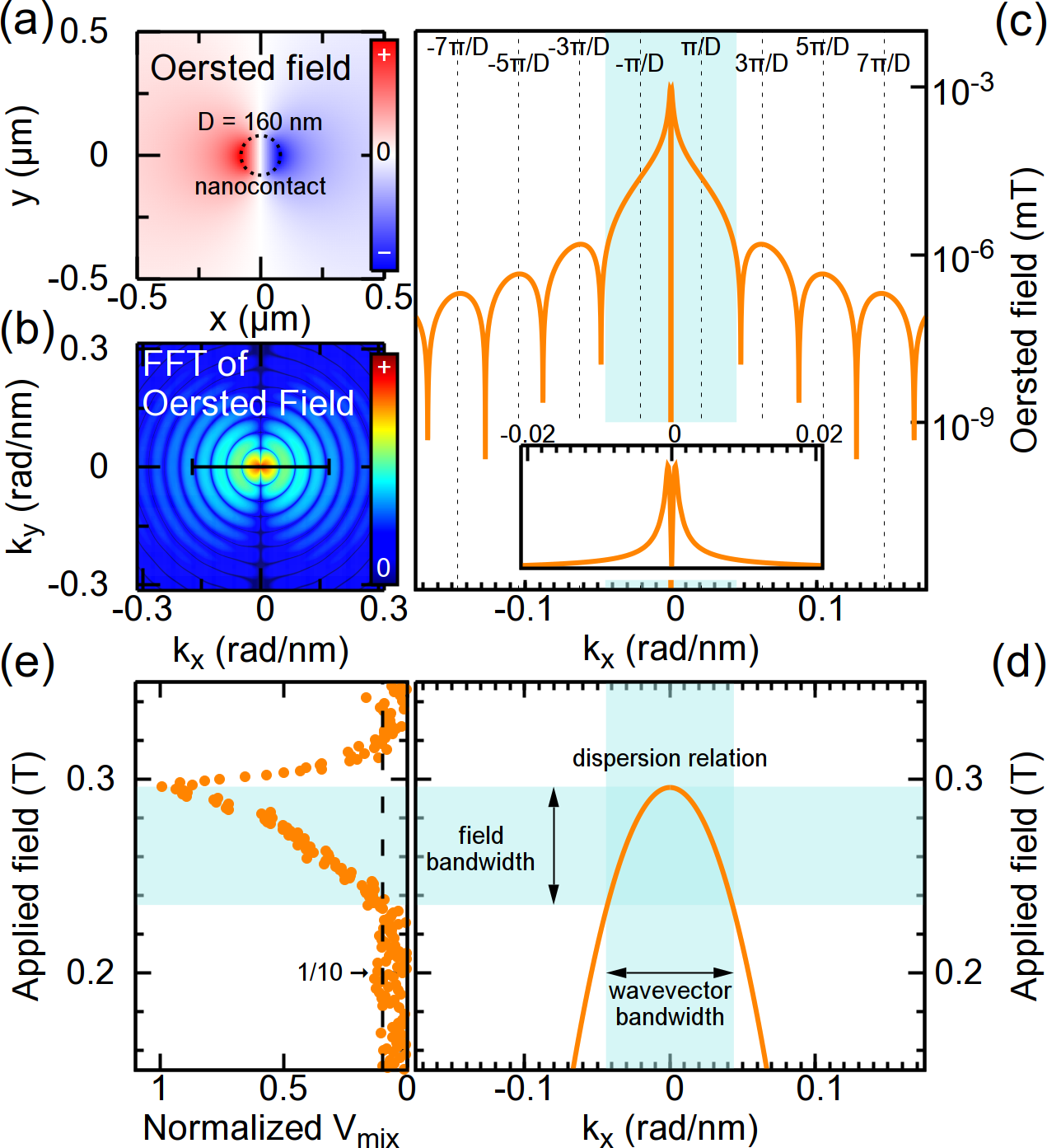} 
	\caption{(a) The spatial distribution and (b) the corresponding 2D FFT of the out-of-plane component of the Oersted field by the D = 160 nm diameter NC. (c) The Oersted field as the function of the wavevector component along the saturation direction is shown for the D = 160 nm diameter NC. The inset shows a zoom-in of the small wavevector part of the spectrum. (d) The dispersion relation of the exchange dominated SWs. (e) The experimentally acquired magnetization dynamics spectrum using nanocontact of D = 160 nm diameter.
    \label{fig:7}}
\end{figure}

Due to the collinear free and fixed layers we do not expect any significant contribution from the ST to the observed magnetization dynamics, which is confirmed by comparing micromagnetic simulations performed with and without ST included. Correspondingly, for the in-plane applied magnetic fields in the NC geometry with no \emph{dc} bias currents applied, the \emph{rf} Oersted field is the primary excitation mechanism responsible for the observed dynamics.

In a linear approximation the response of the system is essentially determined by the spectrum of the excitation, which in our case is provided by the \emph{rf} Oersted field. If the excitation has a finite amplitude at some point of reciprocal space (\emph{i.e.} at the given frequency and wavevector), then, if allowed, the corresponding magnon will be excited. The spatial profile of the Oersted field and its spectrum for the $D = 160$ nm NC are shown in Fig.~\ref{fig:7}(a) and (b), respectively. We can identify both local and global anti-symmetries with respect to the NC center with corresponding periods determined by twice the NC diameter and mesa width, $L$, respectively. Since both spatial components are naturally confined to their unit periods, the linewidth of the corresponding excitation peak is finite. Therefore the Oersted field most efficiently couples to the SW bands having widths of $2\pi/L$ and $\pi/D$ corresponding to the wavevectors of $2\pi/L$ and $n\pi/D$, respectively, where $n=1,3,...$ (see Fig.~\ref{fig:7}(c) and its inset). As the NC diameter decreases, the position and width of the former band stays constant, while the latter one shifts towards lower resonance fields and increases its width, leading to the observed extension of the tail in the excitation spectrum.

It is important to mention that the circular NC cannot effectively couple to the uniform FMR. Instead, the main peak observed in the experiments and simulations corresponds to the anti-symmetric mode with $k=2\pi/L$. However, due to the vanishing magnetostatic dispersion, its resonant field is virtually indistinguishable from the uniform $k=0$ FMR mode.

Considering a typical FMR experiment where the excitation frequency is fixed, according to Eq.~(\ref{Equation:3}) the wavevector of the generated propagating SW is ultimately determined by the value of the applied magnetic field. As the field is swept towards zero past the dominant FMR resonance, the NC continuously excites propagating SWs of increasing wavevectors. Since the excitation amplitude drops rapidly for the low values of the applied field (\emph{i.e.} for short wavelength SWs), the detected magnetic signal vanishes accordingly, leading to the appearance of the low-field tail, see Fig.~\ref{fig:7}(c) and (e).

By assuming that the extent of the tail is estimated at 1/10 of the its peak amplitude, we can project the corresponding experimentally observed applied magnetic field to the cutoff wavevector of the excitation spectrum as schematically demonstrated by the shaded rectangles in Fig.~\ref{fig:7}(c), (d) and (e). This gives us the cut-off wavevectors (in units of $\pi/D^{\prime}$) of 1.93, 1.95 and 2.34 for the NCs of 90, 160 and 240 nm nominal diameters, respectively. Since these values fall roughly inside the first two fundamental SW bands attributed to $k=2\pi/L$ and $k=\pi/D$, the two-peak scheme used to fit the experimental data is fully justified.

It should be noted that the micromagnetic simulations do not reproduce the shoulder as it is observed experimentally for all the NC diameters. According to our model, the shoulder should be inherited from the excitation spectrum. Perhaps the approximation we used to calculate the Oersted field, an infinite wire, is not sufficient to bring out this feature. Nevertheless, this does not change the interpretation of the results and conclusions of the present study.

Finally, the NC size dependence of the FMR and SWR inhomogeneous broadenings shown in Fig.~\ref{fig:3} can be well understood by assuming that it is inherited from the linewidth of the corresponding excitation peaks. For the SWR mode, the expected extrinsic contribution to the magnonic linewidth is 96 mT, 43 mT and 15 mT for the NC diameters of 82 nm, 122 nm and 205 nm, respectively, in excellent agreement with the fitted values. In contrast, for the FMR the contribution is vanishing and should be virtually independent of the NC size. However, if the NC had shape imperfections, the corresponding irregularities in the Oersted field profile should broaden the excitation peaks and, eventually, the FMR and SWR. As we typically observe a less perfect NC for smaller diameters, the inhomogeneous broadening of FMR should increase accordingly, consistent with the experimental data.

\section{Conclusions}
In conclusion, using homodyne based measurement techniques we provide an in-depth study of the magnetodynamics in a quasi-confined system, namely a NC patterned on an extended pseudo spin valve film stack. The observed spectra are highly asymmetric and cannot be explained by a single resonance mode, as has been done in the past \cite{Tsoi2011, Tsoi2013}. Instead, each spectra is fit by a combination of two Lorentzians from which we can extract the FMR mode resonance field and linewidth. The secondary mode corresponds to the generation of exchange-dominated spin waves with a wavevector inversely proportional to the NC diameter. The results are reproduced by the micromagnetic simulations that show the \emph{rf} Oersted field generated by the injected \emph{rf} current is the dominant excitation mechanism of the observed magnetization dynamics. We thereby demonstrate experimentally a highly tunable point source of the propagating SW with the wavevectors limited only by the resolution of the fabrication process used. This is of the paramount importance for the applications of sub-THz and THz magnonics and spintronics.

\section{Acknowledgments}
We would like to thank Pranaba Muduli and Mojtaba Ranjbar for useful discussions.  This work was supported by the European Commission FP7-ICT-2011-contract No. 317950 ``MOSAIC''. It was also supported by the European Research Council (ERC) under the European Community\textsc{\char13}s Seventh Framework Programme (FP/2007-2013)/ERC Grant 307144 ``MUSTANG''. Support from the Swedish Research Council (VR), the Swedish Foundation for Strategic Research (SSF), the G\"{o}ran Gustafsson Foundation, and the Knut and Alice Wallenberg Foundation is also gratefully acknowledged.  MD would like to thank The Wenner-Gren Foundation. 

\section{References}

\bibliographystyle{aipnum4-1}

\end{document}